\documentclass[aps,pre,reprint,showpacs]{revtex4-1}
\usepackage{graphicx}
\usepackage{dcolumn}
\usepackage{bm}
\usepackage{color}
\usepackage{ulem}
\usepackage{xfrac}
\usepackage{float}
\usepackage{hyperref}

\newcommand{\mr}[1]{\mathrm{#1}}

\begin{document}

\title{DNA-Programmed Mesoscopic Architecture}
\author{Jonathan D. Halverson}
\author{Alexei V. Tkachenko}
\affiliation{Center for Functional Nanomaterials, Brookhaven National
Laboratory, Upton, New York 11973}
\email{Electronic address: oleksiyt@bnl.gov}
\date{\today}

\begin{abstract}
We study the problem of the self-assembly of nanoparticles (NPs) into finite
mesoscopic structures with a programmed local morphology and complex
overall shape. Our proposed building blocks are NPs
directionally-functionalized with DNA. The
combination of directionality and selectivity of interactions allows one
to avoid unwanted metastable configurations which have been shown to
lead to slow self-assembly kinetics even in much simpler systems.
With numerical simulations, we show that a variety of target
mesoscopic objects can be designed and self-assembled in near perfect
yield. They include cubes, pyramids, boxes and even an Empire State
Building model. We summarize our findings with a set of design strategies that
lead to the successful self-assembly of a wide range of mesostructures.
\end{abstract}

\maketitle

\section{Introduction}

Impressive progress has been made recently in using the molecular
recognition properties of DNA to direct the self-assembly of
nanoparticles~\cite{mirkin1996,alivisatos1996} into a variety of periodic
nanostructures~\cite{nykypanchuk2008,park2008,macfarlane2011}. Many
experimental~\cite{crocker2006,leunissen2009,crocker2009,sinno2011,crocker2011}
and theoretical~\cite{tkachenko2002,starr2010,atv_frenkel2011,travesset2011,starr2011,stefano2012} studies
have aided our understanding of these systems and helped in formulating
design rules. However, the NP-DNA superlattices
that have been the focus of most of the research to date,
represent only a small subset of the plausible morphologies for this class
of systems. One may expect that DNA hybridization would enable one to
design arbitrary nanoparticle clusters with a prescribed overall
architecture. In this work, we explore a methodology for
``programming'' such NP-DNA mesoscopic structures, and test
it with computer simulations. We demonstrate that the slow kinetics and errors
typically associated with unwanted metastable states can be completely
avoided for nanoparticles which are \textit{directionally functionalized}
with DNA.

\begin{figure}[tbp]
\centering \includegraphics[width=7.6cm]{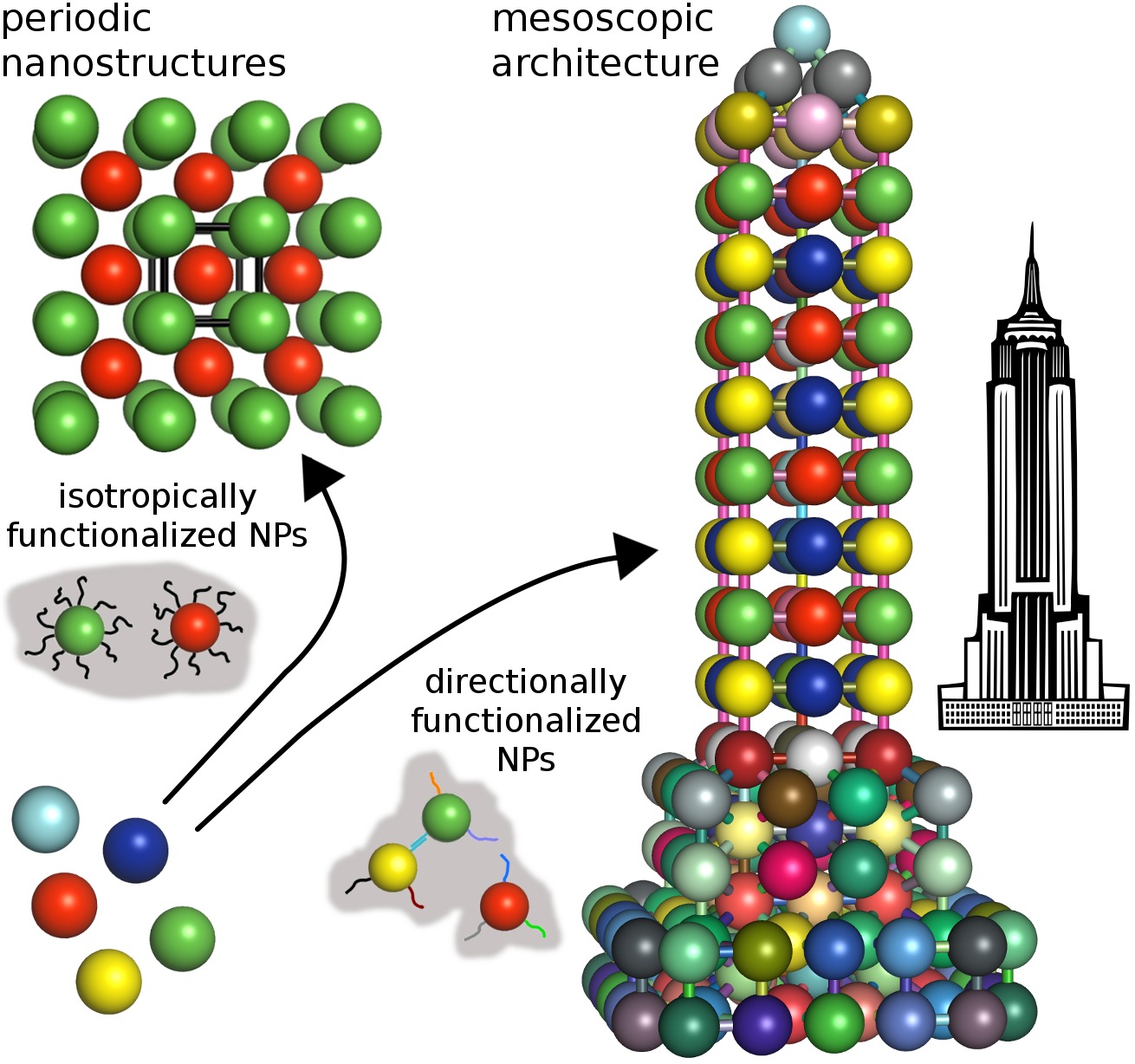}
\caption{(Color online) Nanoparticles isotropically functionalized with
complementary single-stranded DNA self-assemble into periodic nanostructures
like a body centered cubic crystal. When the
nanoparticles are directionally functionalized by attaching DNA strands at
specific locations, the particles can be designed to self-assemble into
finite-size mesoscopic architectures like the Empire State Building.}
\label{overall}
\end{figure}

Conceptually, we are looking for the solution of the inverse problem in
self-assembly: To identify the set of constituent particles that will form
the desired structure in a robust, error-free manner. In a finite set of
colloids isotropically functionalized with DNA, one can design the pairwise
DNA-mediated interactions so that a particular target cluster will be its
ground state~\cite{licata2008,meng2010,brenner2011}. However, it has been
shown both theoretically and experimentally that the space of metastable
configurations in such a system grows rapidly with the cluster size, making
the overall kinetics very slow. Furthermore, in order to program a
structure containing $n_0$ particles one would typically need to use $n_0$
distinct DNA sequences to encode each individual particle differently, and isolate a
complete set of particles prior to the self-assembly. In the theoretical paper
that precedes this work~\cite{tkachenko2011}, we have argued that these
problems are avoided by the use of directionally-functionalized
nanoparticles (dfNPs) (see Fig.~\ref{overall}). It has been argued
that the target structure is kinetically preferred if the partially built
mesostructure is sufficiently rigid to avoid undesired bonds. We have also
estimated the upper bound of the number of DNA sequences needed to program a
particular structure. In this work, we use computer simulations to move from
that abstract theoretical foundation towards more specific and practical
realizations.

The dfNPs are more complex than regular DNA-functionalized particles due to
the fact that single-stranded DNA (ssDNA) with given sequences are attached
at well-defined positions rather than randomly. Recent experiments indicate 
that achieving a greater control over DNA functionalization and eventually
building arbitrary  dfNPs is within reach. For instance, Suzuki et al.~\cite{suzuki2009} have
made particles with two distinct ssDNA attached at a
well-defined separation. Kim et al.~\cite{kim2011} used electrostatic
repulsion to form dfNPs with up to 6 ssDNA, and colloidal particles with
controlled DNA surface patches have also been synthesized~\cite{wang_directional}. Theoretically,
we proposed several pathways to
self-assemble dfNPs~\cite{licata2008,tkachenko2011}, inspired by advances in
DNA nanotechnology~\cite{winfree1998,barish2009,rothemund2006}. These
building blocks, in addition to the mutual recognition of
the neighboring particles, provide control over valence (number of bonds per
particle) and, to a large extent, bond directionality. In other words, dfNPs
combine the advantages of two major classes of building blocks proposed for
programmable self-assembly: high selectivity of binding typical for
isotropic DNA-functionalized particles and the directionality of
interactions similar to that of patchy colloids~\cite{glotzer2007}.

One can also make a parallel between our approach and the booming field of
DNA nanotechnology. Similar to DNA origami~\cite{rothemund2006} and most
recent DNA Lego~\cite{ke2012} approaches, our objective is to create a
mesoscopic 3D object with predetermined architecture, and similar to DNA
tiling and algorithmic self-assembly \cite{winfree1998,barish2009,park_tiles}, we
rely on selective directional interactions to program the structure. 
However, in our case NPs of an arbitrary type are naturally
incorporated into the design, which greatly expands the range of potential
applications.

\section{Model and Methodology}

The central idea of our approach is to encode bonds between neighboring NPs in the
target state with a pair of complementary ssDNA, each grafted to its respective particle.
This design procedure maps the
desired mesostructure onto a set of dfNPs. Upon
hybridization of the ssDNA pairs, the rigid double-stranded DNA (\mbox{dsDNA}) bonds
are formed. If the scheme is successful a collection of
identical programmed mesostructures will assemble.
Here we engineer a number of increasingly complex
mesostructures, and use Brownian dynamics simulations to verify that the
mesostructures self-assemble from an initially disordered solution of dfNPs.

We have in mind NPs of size 5--50 nm in solution.
In the simulations, particles undergo translational and rotational Brownian motion subject to
a pairwise excluded-volume potential
$U_{\mathrm{ev}}(r)=A\exp[-B(r-2a)]$, where $r$ is the separation distance and $a$ is the particle radius.
The DNA bonds are modeled by a harmonic potential $U_{\mathrm{bond}}(l)=k_s(l-l_0)^2/2$ with $l_0=a/2$.
The particle volume fraction is $\phi =(4/3)\pi a^{3}N/L^{3} \sim 10^{-3}$, where $N$ is the number
of particles and $L$ is the side of the cubic simulation box.
Periodic boundary conditions were used.

The forming and breaking of dsDNA bonds was treated using a stochastic scheme based on a local equilibrium model.
For a binary system where each NP is grafted with a single DNA chain of sequence $\mathrm{A}$ or its complement $\mathrm{A}'$,
the dissociation and association rate constants are related by

\begin{equation}
k_d = k_a\frac{C}{4}e^{x},
\label{rate_constants}
\end{equation}

\noindent
where $x \equiv \Delta S(T - T^{\ast})/RT$ and $C=N/L^3$. In the definition of $x$,
$\Delta S$ is the entropy of
hybridization of free DNA, $R$ is the universal gas constant and $T$ is the absolute temperature.
Eq.~\ref{rate_constants} has been written such that the equilibrium fraction of dimers is $1/2$ at $T=T^{\ast}$,
where $T^{\ast}$ is approximately equal to the melting temperature of free dsDNA with the sequence $\mathrm{AA}'$
at concentration C. The control parameter $x$ characterizes the deviation from $T^{\ast}$ and in
general it can be different for each dsDNA bond in the mesostructure.

Bond forming and breaking was attempted at each time
step of the simulation after the positions and orientations of the particles
were updated.
Our approach uses an implicit model of DNA where only the surface positions or site locations of the ssDNA are needed.
Particle pairs with complementary sites that are not bonded and separated by less than $l_{0}$ are added to a list of possible site-site bonds.
The number of bonds to actually form from this list (randomly chosen) is drawn from a Poisson distribution with
mean $n_a \kappa \Delta t / V_R$, where $V_R=(2/3)\pi l_{0}^3$, $n_a$ is
the total number of possible bonds and $\Delta t$ is the time window which was taken as the time step.

For the disassociation, the number of bonds to randomly break is drawn from a Poisson distribution
with mean $(1/4)\alpha k_a C \Delta t$, where $\alpha=\sum_k e^{x_k}n_k$ and $n_k$ is the number
of existing bonds of type $k$.
In our procedure the probability of breaking a bond of type $k$ is proportional to $e^{x_k} n_k/\alpha$.
In the simulations we use $\kappa\tau/a^3=10$ which corresponds to $k_a\tau/a^3=0.42\pm0.03$ as explained below.
The translational diffusion
time is $\tau =a^{2}/6D$ with $D$ given by the Stokes-Einstein
relation $D=k_{B}T/6\pi \eta a$, where $k_B$ is the Boltzmann constant and $\eta$ is the solvent viscosity.

\begin{figure}[htpb]
\centering \includegraphics[width=8cm]{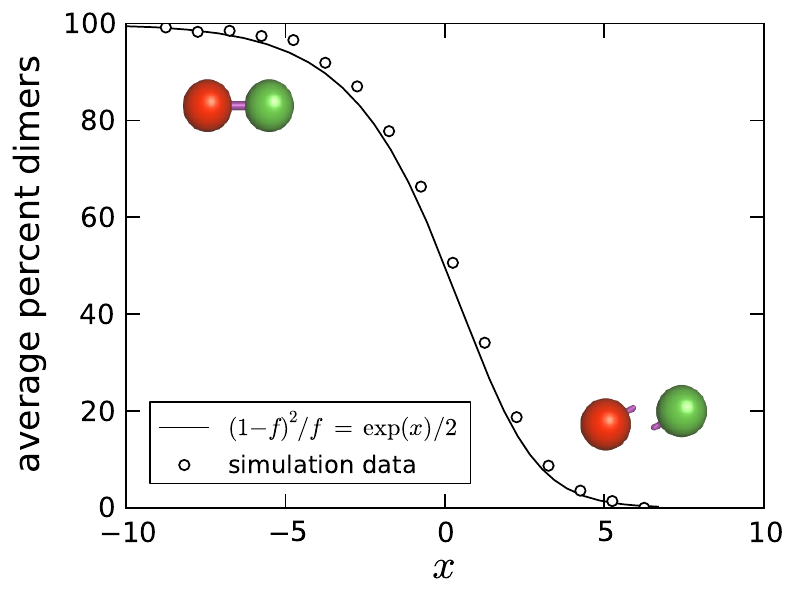}
\caption{(Color online) Melting curve for dimers with $N=96$ and $\phi=0.0025$. The solid line
is the theoretically expected melting curve given by Eq.~\ref{theory_dimers}.}
\label{dimers_data}
\end{figure}

The melting curve for dimers as determined by simulation is shown in Fig.~\ref{dimers_data}.
For $x=5$ there are a very small number of dimers. This value
corresponds to roughly the melting temperature of the dsDNA bond. As $x$ decreases the number of dimers increases
until around $x=-10$ where all possible bonds have formed. It is important to realize that for a
given value of $x$ these systems enter a dynamic equilibrium where
bonds are broken and reformed.
The simulation data is qualitatively consistent with experimentally obtained
curves~\cite{leunissen2009}. One may derive an analytical expression to describe the data.
The time evolution of the concentration of dimers $c_{AA'}$ is given by

\begin{equation}
\dot{c}_{AA'}=k_a c_A c_{A'} - k_d c_{AA'}.
\label{time_ev}
\end{equation}

\noindent
With $c_{AA'}=fC/2$, where $f$ denotes the fraction of dimers, at steady state Eq.~\ref{time_ev} becomes

\begin{equation}
\frac{(1-f)^2}{f}=\frac{e^x}{2}.
\label{theory_dimers}
\end{equation}

\noindent
Good agreement is found between the simulation data and Eq.~\ref{theory_dimers} as shown in Fig.~\ref{dimers_data}.
Note that the melting curve in Fig.~\ref{dimers_data} is for dimers at a given value of $\phi$ and it does not
necessarily apply to the systems described below.
Additional details of our model and simulation methodology are given in the Supplemental Material (SM)~\cite{supmat}.

\section{Results}

\begin{figure}[htpb]
\centering \includegraphics[width=8cm]{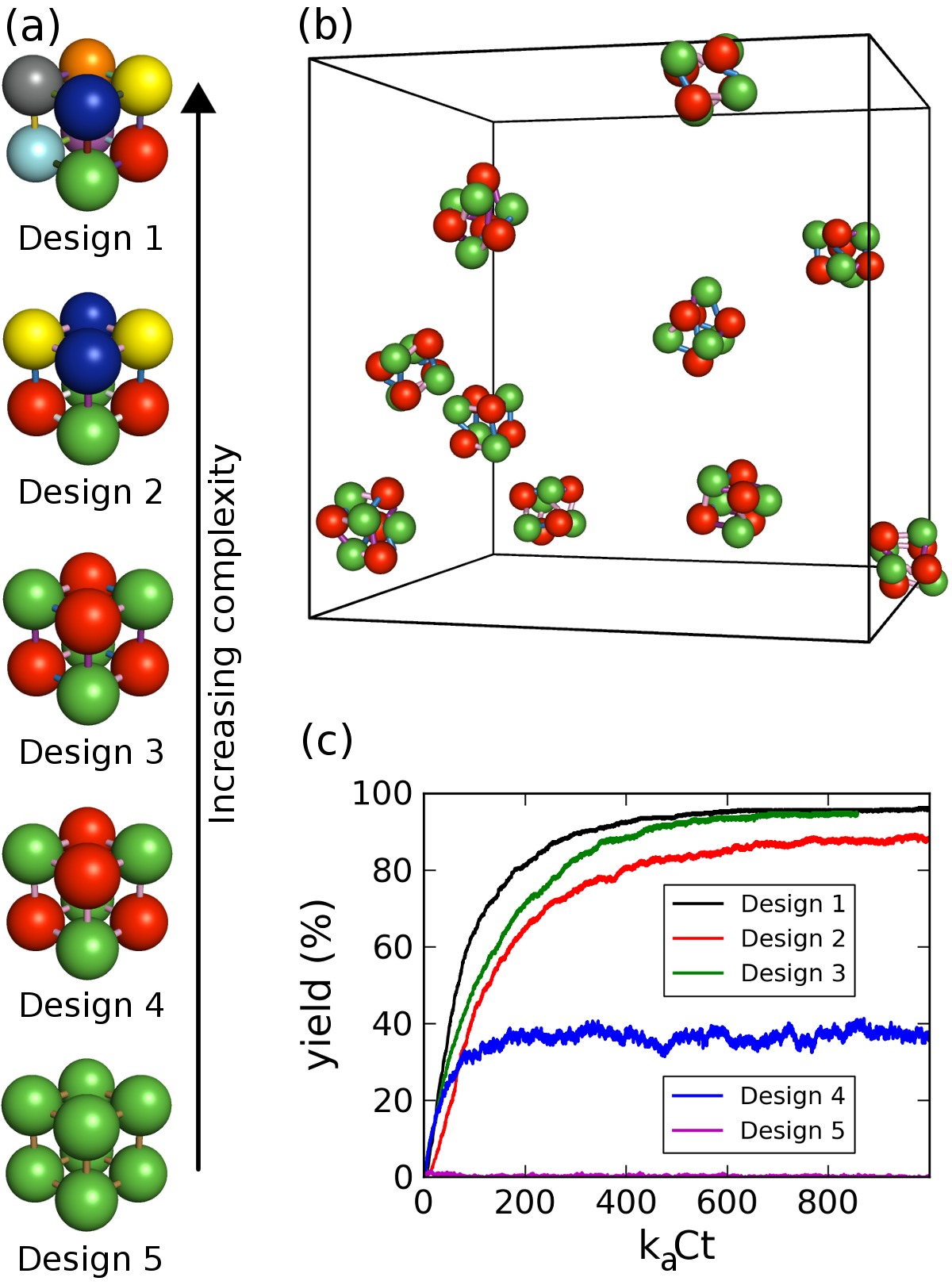}
\caption{(Color online) (a) Five designs with the NPs arranged on the
vertices of a cube. For each design the unique complementary ssDNA pairs or
bonds are drawn with a unique color and particles of the same color are
identical.
For Designs 1--5, the number of unique bonds/particle types are 12/8, 4/4, 3/2, 1/2, 1/1, respectively.
Note that Designs 3 and 4 have
the same number of particle types but differ in the number of unique bonds.
(b) Final configuration of the system with Design 3. This simulation was conducted at $\phi
=10^{-2}$ and $N=80$ for illustrative purposes. (c) Percent yield of cubes versus time
for the five designs with $\phi=10^{-3}$ and $N=200$. The data has been averaged over
24 independent systems for each design. At $k_aCt=800$, the yield in increasing
order is Design 5, 4, 2, 3, 1 with values of roughly $0, 40, 90, 95, 95\%$, respectively.}
\label{cubes_fig}
\end{figure}

Consider a simple mesostructure where
the particles are arranged on the vertices of a cube.
The most complicated
and robust design would be to encode each bond with a unique pair of
complementary DNA strands leading to 12 complementary pairs and 8 particle
types (Design 1 in Fig.~\ref{cubes_fig}(a)). We can use the underlying symmetry of
the target structure to reduce the number of unique DNA sequences and particle
types. This process resulted in 5 different designs with gradually
reducing complexity as shown in Fig.~\ref{cubes_fig}(a).

Simulations have been conducted to study the self-assembly of each of the
five cube designs. At initialization, the particles are assigned random
positions and orientations.
The $x$ value for each bond in each design was the same with
$x=-0.25,~0.625,~0.35,~2.25,~2.25$ for Designs 1--5, respectively.
The value of $x$ varies between
designs because the concentration of bond types is different for each design.
The percent yield is shown
in Fig.~\ref{cubes_fig}, where the yield is defined as the number
of mesostructures perfectly formed, with all particles and bonds in place, divided by the
total number of expected mesostructures. Designs 1--3 lead to cubes in high yield.
While Design 3 has only 3 unique
complementary pairs it leads to the same yield and fast kinetics as the
design with 12 pairs (see Video 1 of SM~\cite{supmat}). Design 4 is
produced in a yield of just less than 40\%, and Design 5 leads to the formation
of tetrahedra and 6-particle clusters in high yield with essentially no cubes.
The standard deviation in the final yield
was less than 3\% for all designs except Design 4 which was approximately 7\%.
From the experimental point of view, Design 3 is optimal because it requires
the fewest number of unique DNA sequences to produce the target cluster
quickly and in near perfect yield.

\begin{figure*}[!htp]
\centering
\includegraphics[width=16cm]{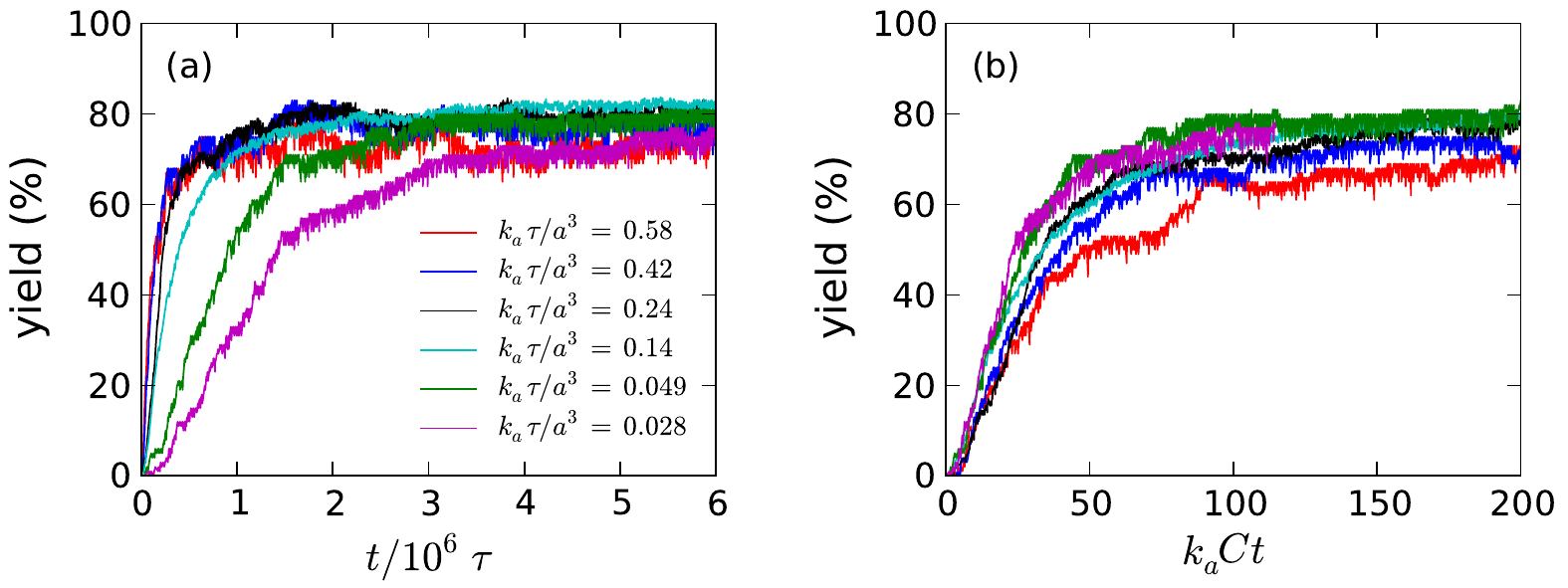}
\caption{(Color online) Effect of hybridization rate constant on cube systems with Design 3
and $N=160$, $x=1.25$ and $\phi=0.0025$.
(a) Percent yield versus time. At early times the yield increases monotonically with increasing values of $k_a$.
(b) Percent yield versus time normalized by $1/k_aC$. The data sets are found to collapse with the $k_a\tau/a^3=0.58$ case somewhat off the master curve.}
\label{cubes_ka}
\end{figure*}

We have investigated the effect of the DNA hybridization rate on the
self-assembly of cubes with Design 3.
For a given choice of $\kappa$ there is a corresponding value of $k_a$.
To determine these values we conducted simulations of the self-assembly of dimers with $N=512$.
By choosing $x=-15$ to prevent disassociation, $k_a$ can be obtained as the initial slope
of a plot of $1/c_A$ versus $t$.
For instance, with $\kappa\tau/a^3=0.1,~1,~10$, we find $k_a\tau/a^3=0.0069 \pm 0.0003,~0.049 \pm 0.002,~0.42 \pm 0.03$, respectively.
The yield of cubes as a function of time
for six values of $k_a$ is shown in
Fig.~\ref{cubes_ka}(a).
All cases give an
equilibrium yield of roughly 65\% with the time required to
reach equilibrium decreasing with increasing $k_a$. In Fig.~\ref{cubes_ka}b
the yield is plotted against $k_aCt$. While $k_a$ spans a wide range, the
data sets are found to mostly collapse onto a master curve indicating that the
characteristic time scale for these systems is $1/k_aC$. The exception to this behavior is the
curve with the largest value of $k_a$.
In the simulations we use $\kappa\tau/a^3=10$, and its corresponding $k_a$ value of $0.42~a^3/\tau$, because the structures
form fast with this choice and the overall yield is the same as for smaller values.

One expects $k_a$ to be less than
but within an order of magnitude of the association rate constant of free DNA $k_a^{\mr{(free)}}$.
If an experimental value were used for $k_a$ one would
expect the same results but the simulations would have to be run for much longer.
For instance, Tsuruoka et al.~\cite{tsuruoka1996}
reported a value of $k_a^{\mr{(free)}}$ of
roughly $10^5$~s$^{-1}$M$^{-1}$ which corresponds to $k_a\tau/a^3 \sim 10^{-3}$.
Chen et al.~\cite{chen2007} report for complementary strands with 21 bases a value of $10^4$~s$^{-1}$M$^{-1}$.
Hence, in real experiments, a rough estimate of the characteristic time scale
is $1/k_aC \sim 25$~s, which indicates that the cubes form in roughly 4 hours.

\begin{figure*}[htbp]
\centering \includegraphics[width=17cm]{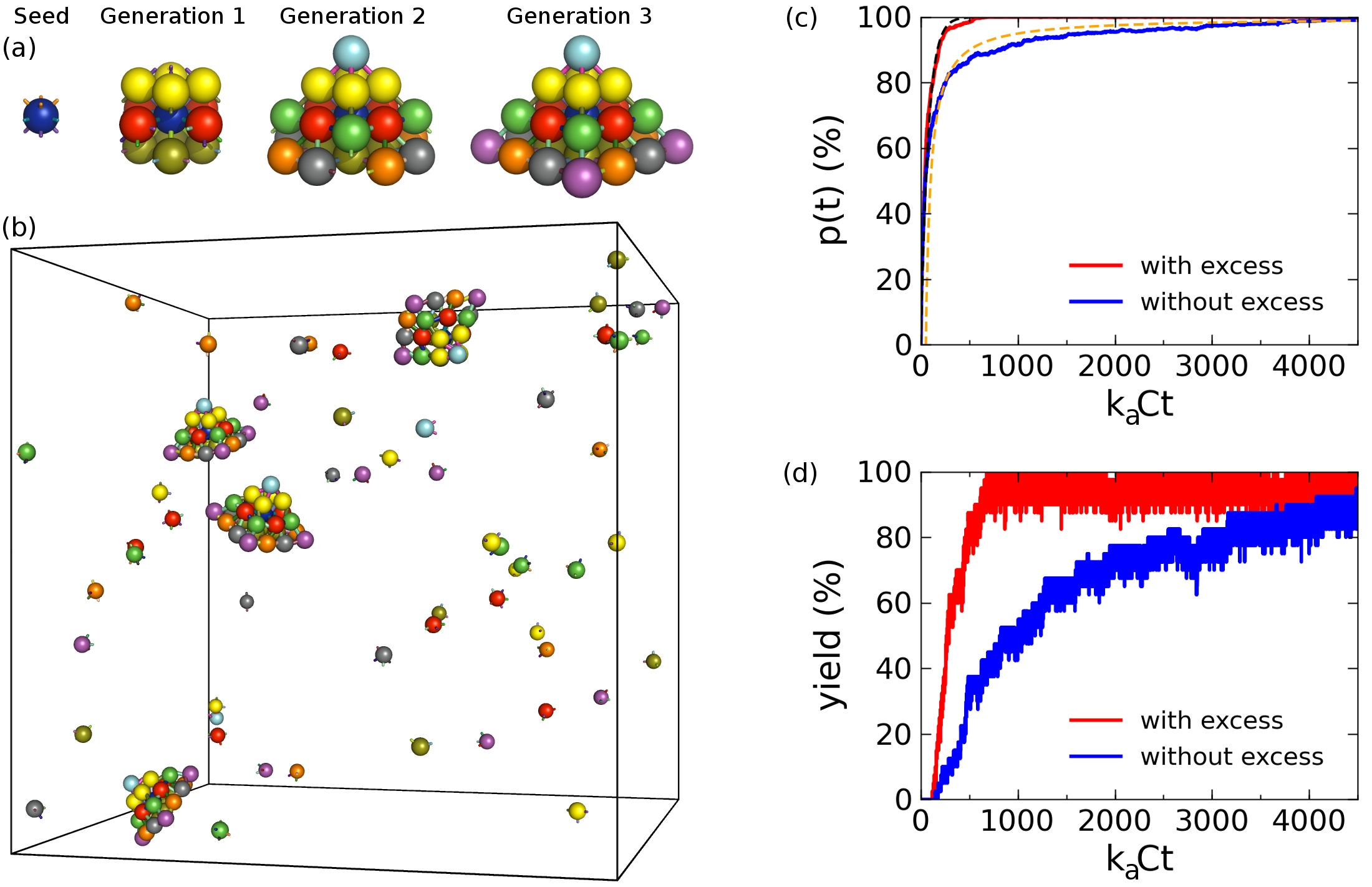}
\caption{(Color online) (a) Pyramids can be formed in a
generation-by-generation manner by using a central seed particle. While the
pyramid is composed of 60 bonds and $n_0=30$ particles, our encoding scheme needs
only 15 unique bonds and 9 particle types.
(b) Pyramids self-assemble from an initially disordered solution of dfNPs
with excess particles ($\phi=0.0025,~N=178$). (c) $p(t)$ and (d) yield
for the pyramids with excess particles ($N=178$, red (light gray) line) and without ($N=120$, blue (dark gray) line). The data is averaged over
10 independent simulations. The dashed black and orange (gray) lines in (c) are fits to the data (see SM).}
\label{pyramids_fig}
\end{figure*}

To fabricate structures more complicated than cubes it becomes important to
employ a strategy that controls both the process of nucleation and growth.
Ideally, one wants to operate in the regime where the number of nucleation
sites or dynamically-growing clusters matches the total number of expected mesostructures,
with all other particles remaining unbound until joining a cluster. Such
conditions can very nearly be realized by forming the mesostructure in a
generation-by-generation manner as illustrated for a pyramid in Fig.~\ref{pyramids_fig}(a).
The success of this design strategy, which is described in detail below, is
demonstrated in Fig.~\ref{pyramids_fig}(b) where the pyramids are shown to self-assemble in
perfect yield (see Video 2 of SM~\cite{supmat}). Simulations have been conducted when
the system composition corresponds to exactly four pyramids and when excess particles (not seed) are present.
Aside from the immediate application of symmetry, the design of the pyramid
was arrived at through an iterative process where the number of bonds and $x$-values were refined
until the structures formed quickly and without errors (see SM for details).
Over the course of this study a general design strategy became apparent.
This strategy, which is applicable to all the mesostructures described
here, is given below.

\begin{figure*}[htbp]
\centering \includegraphics[width=17cm]{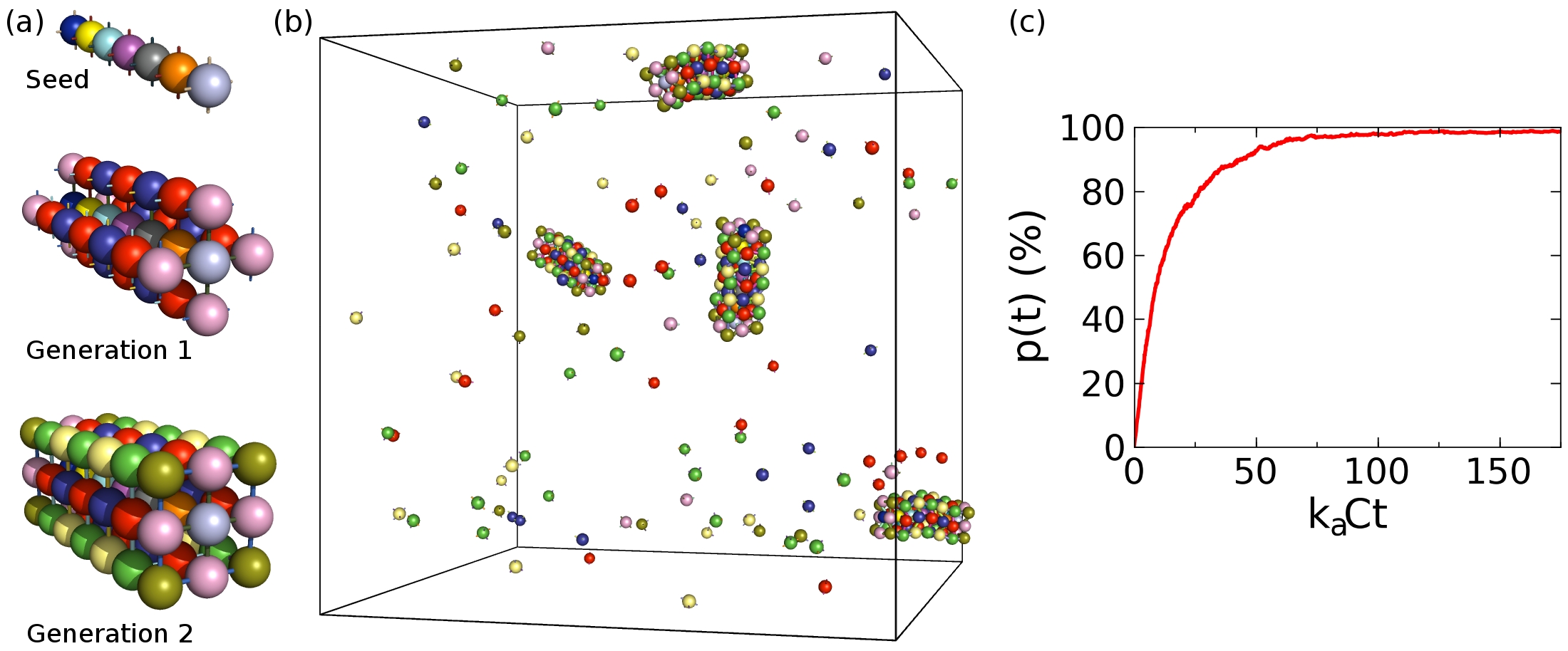}
\caption{(Color online) (a) An orthorhombic box structure self-assembles in
a generation-by-generation manner when templated by a linear seed. The 114
bonds and $n_0=63$ particles of the structure are encoded using only 11 unique
bonds and 13 particle types.
(b) Final configuration of a box
simulation with $\phi=0.0025$ and $N=364$.
(c) $p(t)$ for the box mesostructures
during the low temperature phase averaged over 10 independent systems.}
\label{box_fig}
\end{figure*}

For typical $x$ values, the bonds between the NPs in a fully formed mesostructure are continually breaking and reforming.
For large structures where it
is very likely that at least one bond will be broken at time $t$, our previous definition of the yield
is replaced by
$p(t) =\langle (n(t)-n_s)/(n_0-n_s) \rangle$, where $n(t)$ is the number of particles
in the dynamically-growing cluster, $n_s$ is the number of seed particles,
and the brackets denote an average over
all clusters. Fig.~\ref{pyramids_fig}(c) and (d) show the yield of pyramids using both definitions.
The standard deviation in $p(t)$ at equilibrium is essentially zero for the case with
excess particles and approximately 2\% for the case without.

When the clusters are in an intermediate
state of growth, a simple description of the time evolution
of the concentration of particles of type $i$ that
have yet to join the clusters is

\begin{equation}
\dot{c}_i = -k_a^{(i)} (c_i + c_{\mr{ex}})c_i,
\label{vacancies}
\end{equation}

\noindent
where $c_{\mr{ex}}$ is the concentration of excess particles and $k_a^{(i)} \sim k_a$.
For $c_{\mr{ex}} = 0$ the concentration of vacancies
decays slowly as a power law, $c_i \sim 1/k_a^{(i)} t$, while adding
excess particles leads to sharper exponential relaxation,
$c_i \sim \exp(−k_a^{(i)} c_{\mr{ex}} t)$. Good agreement is found when these
expressions are used to fit the data in Fig.~\ref{pyramids_fig}(c).
In real experiments the pyramids are predicted to form in roughly 1 hour.

While symmetry is helpful in reducing the design complexity, there are many
finite-size objects with a repeating motif which lack symmetry in a strict
mathematical sense. For instance, in the 7-layer box structure shown in
Fig.~\ref{box_fig}(a), the translational symmetry is broken due to its finite size. Does
this imply that each of the layers has to be built out of a distinct set of
particles? In the proposed design, we avoided this by using a linear seed
made of 7 distinct particles. The seed effectively encodes the size and 1D
coordinate along the cluster.
The rest of the box (56 particles) is built around the seed with only 6
distinct particle types.
This means that the length of the structure can be increased with a
minimal increase in the number of particle types. While the
cubes and pyramids form at a single temperature, the linear seed of the box forms
at, for example, room temperature with the remaining particles joining at a
lower temperature ($\Delta x=-5$).
Boxes are produced in near perfect yield as shown in Fig.~\ref{box_fig}(b) and (c).
The standard deviation in the final value of $p(t)$ was 2\%.

When dfNPs are synthesized in the laboratory it is likely that there
will be slight errors in the positions of the DNA strands. To investigate
the effect of these anticipated errors, simulations using Design 3 for the cubes were
carried out with particles that had each of their surface sites randomly
displaced. The sites were randomly rotated along the particle
surface by two angles about orthogonal axes each according to a Gaussian
distribution with mean $0^{\circ}$ and standard deviation $\sigma$.
The yield of cubes was found to be higher than 75\% for values of $\sigma$ up to
$10^\circ$ and then drop to around 60\% for $20^\circ$ and around 20\% for
$30^\circ$. For a particle with a 10~nm diameter a displacement along the surface by
$10^{\circ}$ is an arc length of about 1 nm or 20\% of the radius. However, even with such a
fairly large error the yield is still high.

We have examined the robustness of our results to variations in the excluded
volume parameters ($A$ and $B$), average bond length ($l_0$) and spring
constant ($k_s$). The simulations were conducted at $\phi=0.0025$ with $N=200$
for the cubes with Design 3 and $N=178$ for the pyramids with excess
particles. The baseline values were $Aa/k_BT=50$, $B a=10$, $k_s a^2/k_BT=50$
and $l_0/a=1/2$. We found that the cubes
and the pyramids still form in high yield ($>75$\%) with variations in $A$ up
to $\pm50$\%, in $B$ up to $\pm20$\%, in $k_s$ up to $\pm50$\% and in $l_0$ from
$+40$ to $-20$\%. These values suggest that our findings are quite robust and
not sensitive to the details of the model.

\section{Discussion and Conclusions}

While the early mesostructure designs were done in a trial-and-error
manner, over the course of this work a general
procedure applicable to a wide range of mesostructures emerged.
The key elements to our design strategy for dfNP systems are summarized below:

I. \textit{Symmetry} -- The complexity of the system can be reduced by taking
advantage of symmetry, subject to two limitations: (\textit{i})
neighbors have to be of different types and (\textit{ii}) all bonds for a given particle
should be encoded by different sequences. While the use of this simple rule is best illustrated
by the cubes, all mesostructures were designed by using symmetry to minimize the number
of unique bond types.

II. \textit{Controlled nucleation} -- For the advanced mesostructures,
the number of nucleation sites is controlled by using a seed structure.
For the pyramid this was a single particle while for the box it was a linear chain.
Particles in the seed structure form strong bonds to particles in the first
generation. This can be accomplished experimentally by using DNA sequences
with a high CG content.
The pyramids form at a single temperature. For the box, the dfNPs are designed such that only the seed
forms at a higher temperature (e.g., room temperature) by using a high CG content in the bonds.
When the temperature is lowered by a few degrees the bonds in the outer generations become strong enough to form.

III. \textit{Cooperative binding} -- The pyramid and box mesostructures self-assemble
in a generation-by-generation manner.
Once the first generation has formed, particles in higher generations can easily add to the cluster
because forming multiple bonds with the particles that are
already in place is energetically favorable whereas dimer formation is not.
This can be understood by
considering a green particle in Generation 2 of the pyramid (see Fig.~\ref{pyramids_fig}(a)).
When an unbound green particle encounters either an unbound
red or yellow particle it can form at most one bond which will break at some
later time because bonds are not permanent as indicated by Eq.~\ref{rate_constants}. However, when
Generation 1 is in place and a green particle
bonds to either a red or yellow particle it is very likely to form two
additional bonds which will keep it in place.
Note that generation-by-generation growth of the cluster occurs locally. That
is, it is not necessary for all the particles in the first
generation to be in place before a particle in the second generation can join.

\begin{figure}[htbp]
\centering \includegraphics[width=8cm]{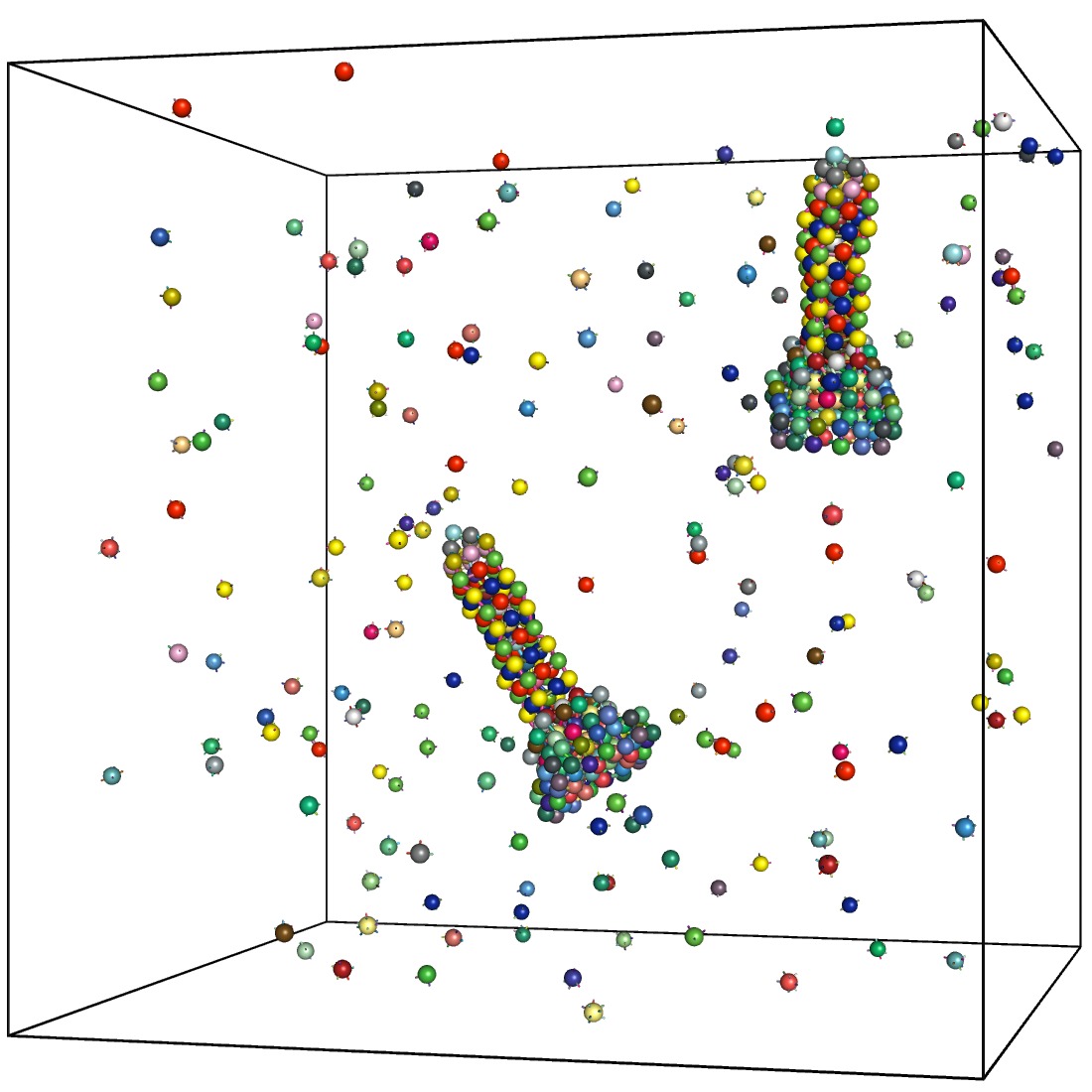}
\caption{(Color online) Final configuration of the Empire State Building
system with excess particles ($\phi=0.0025$, $N=671$).}
\label{esb_fig}
\end{figure}

We have demonstrated that dfNPs can be designed to self-assemble into a
variety of finite mesoscopic structures, each with a programmed local morphology
and complex overall shape. Importantly, this approach avoids
unwanted metastable configurations during the self-assembly process which
allows the mesostructures to form on a time scale that is practical
for nanomanufacturing.
While additional design strategies may be needed as more advanced
mesostructures are sought, the
approach presented here has proven to be sufficient for a wide range of structures.
To illustrate this, we have designed and demonstrated
the successful self-assembly of an Empire State
Building model composed of 210 dfNPs (51 types) as shown in
Fig.~\ref{esb_fig} (see Video 3 of SM~\cite{supmat}). While
it is unlikely that such a structure will be
realized experimentally in the near future, it serves as a remarkable
demonstration of the potential of NP-DNA self-assembly with directional bonding.
The ability to produce objects with complicated programmed
architecture, in a parallel and nearly error-free manner, can eventually
lead to a great variety of practically important applications in the fields
of metamaterials, nanoplasmonics~\cite{fan2010} and nanomedicine.

This work benefited from discussions with O. Gang, M. Hybertsen, W. Sherman,
P. Chaikin, K.-T. Wu and G. S. Grest. Research carried out in whole at the
Center for Functional Nanomaterials, Brookhaven National Laboratory, which
is supported by the U.S. Department of Energy, Office of Basic Energy
Sciences, under Contract No. DE-AC02-98CH10886.

%\bibliography{../letter/dna_assembly.bib}

%merlin.mbs apsrev4-1.bst 2010-07-25 4.21a (PWD, AO, DPC) hacked
%Control: key (0)
%Control: author (8) initials jnrlst
%Control: editor formatted (1) identically to author
%Control: production of article title (-1) disabled
%Control: page (0) single
%Control: year (1) truncated
%Control: production of eprint (0) enabled
%

\end{document}